\documentclass[]{spie}  

\usepackage{float}
\usepackage{amsmath,amsfonts,amssymb}
\usepackage{graphicx}
\usepackage[colorlinks=true, allcolors=blue]{hyperref}

\title{Empirical verification of principal mode orthogonality and relative phase calibration in photonic lanterns}

\author[a,b,c,*]{Adam K. Taras}
\author[a,b]{Barnaby R. M. Norris}
\author[a,b]{Christopher Betters}
\author[a,b]{Daniel S. Dahl}
\author[a,b]{Nathan Long}
\author[a,b]{Andrew Ross-Adams}
\author[a,b]{Peter G. Tuthill}
\author[a,b]{Jin Wei}
\author[a,b]{Sergio Leon-Saval}

\affil[a]{Sydney Institute for Astronomy, School of Physics, The University of Sydney, NSW 2006, Australia}
\affil[b]{Sydney Astrophotonics Instrumentation Laboratory, School of Physics, The University of Sydney, NSW 2006, Australia}
\affil[c]{Leiden Observatory, Leiden University, PO Box 9513, 2300 RA Leiden, The Netherlands}

\authorinfo{Further author information: (Send correspondence to A.K.T.)\\A.K.T.: E-mail: taras@strw.leidenuniv.nl}

\newcommand{\fib}{fibre} 

\usepackage{acronym}	
\usepackage{glossaries}	
\newacronym{USyd}{USyd}{the University of Sydney}
\newacronym{ANU}{ANU}{Australia National University}
\newacronym{ESO}{ESO}{European Southern Observatory}

\newacronym{NSF}{NSF}{National Science Foundation}
\newacronym{LIEF}{LIEF}{Linkage Infrastructure, Equipment and Facilities}

\newacronym{VLTI}{VLTI}{Very Large Telescope Interferometer}
\newacronym{ATs}{ATs}{auxiliary telescopes}
\newacronym{UTs}{UTs}{unit telescopes}
\newacronym{ELT}{ELT}{Extremely Large Telescope}
\newacronym{JWST}{JWST}{James Webb Space Telescope}
\newacronym{DCT}{LDT}{Lowell Discovery Telescope}
\newacronym{GMT}{GMT}{Giant Magellan Telescope}
\newacronym{LBT}{LBT}{Large Binocular Telescope}

\newacronym{STS}{STS}{six telescope simulator}
\newacronym{GPAO}{GPAO}{GRAVITY+ adaptive optics}

\newacronym{BFGS}{BFGS}{Broyden–Fletcher–Goldfarb–Shanno}

\newacronym{WFS}{WFS}{wavefront sensor}
\newacronym{ADC}{ADC}{atmospheric dispersion corrector}
\newacronym{LDC}{LDC}{longitudinal dispersion corrector}
\newacronym{AO}{AO}{adaptive optics}
\newacronym{SNR}{SNR}{signal to noise ratio}
\newacronym{ADU}{ADU}{arbitrary data units}
\newacronym{RMS}{RMS}{root mean square}

\newacronym{OAP}{OAP}{off-axis paraboloid}
\newacronym{DM}{DM}{deformable mirror}
\newacronym{MEMS}{MEMS}{micro-electromechanical system}
\newacronym{IR}{IR}{infrared}
\newacronym{SLM}{SLM}{spatial light modulator}
\newacronym{SLD}{SLD}{superluminescent diode}

\newacronym{SMF}{SMF}{single-mode \fib{}}
\newacronym{MCF}{MCF}{multicore \fib{}}
\newacronym{MMF}{MMF}{multimode \fib{}}
\newacronym{LP}{LP}{linearly polarised}

\newacronym{CAD}{CAD}{computer aided design}
\newacronym{GUI}{GUI}{graphical user interface}

\begin{document} 
\maketitle

\begin{abstract}
Photonic lanterns efficiently map input spatial modes to single-mode outputs for applications like high angular resolution imaging and nulling interferometry. However, manufacturing limits prevent full control over the device's mode transfer matrix at the design stage, making empirical characterisation essential. In this work we further analyse a dataset of direct measurements of a photonic lantern's principal modes using digital off-axis holography over a 73\,nm range near 1550\,nm. By analysing the electric field directly, we find that the principal modes are significantly more orthogonal than random vectors in a space of the same size, as expected for near-adiabatic devices. We propose metrics for quantifying this effect, noting that mode converters with orthogonal principal modes provide better conditioned inverse solvers. We also simulate additional measurements that characterisation systems could take, where the orthogonality would be leveraged to determine the relative phase between principal modes. 
\end{abstract}


\section{INTRODUCTION}

Photonic lanterns provide a means of converting a superposition of a multimodal basis into a set of single modes. Such a device was originally conceived as a means of applying Bragg gratings -- custom wavelength filters -- to light in a \gls{MMF}~\cite{leon-saval_multimode_2005}. This was used to suppress hydroxyl (OH) lines in astrophotonics applications. Photonic lanterns are low loss~\cite{leon-saval_photonic_2013} and can be made in a variety of ways~\cite{birks_photonic_2015}. While the original concept was focused on building high efficiency devices for spectroscopy, the utility of photonic lanterns as mode converters has only recently been widely explored. The most mature developments have been in wavefront sensing: from proof-of-concept~\cite{norris_all-photonic_2020}, theoretical developments with simulations~\cite{lin_focal-plane_2022, lin2023focal} to real-time demonstrations on large telescopes~\cite{lin_real-time_2023}, but other applications such as super-resolution imaging\cite{kim2025sky} are also enabling new science.

\begin{figure}[h]
    \centering
    \includegraphics[width=0.99\linewidth]{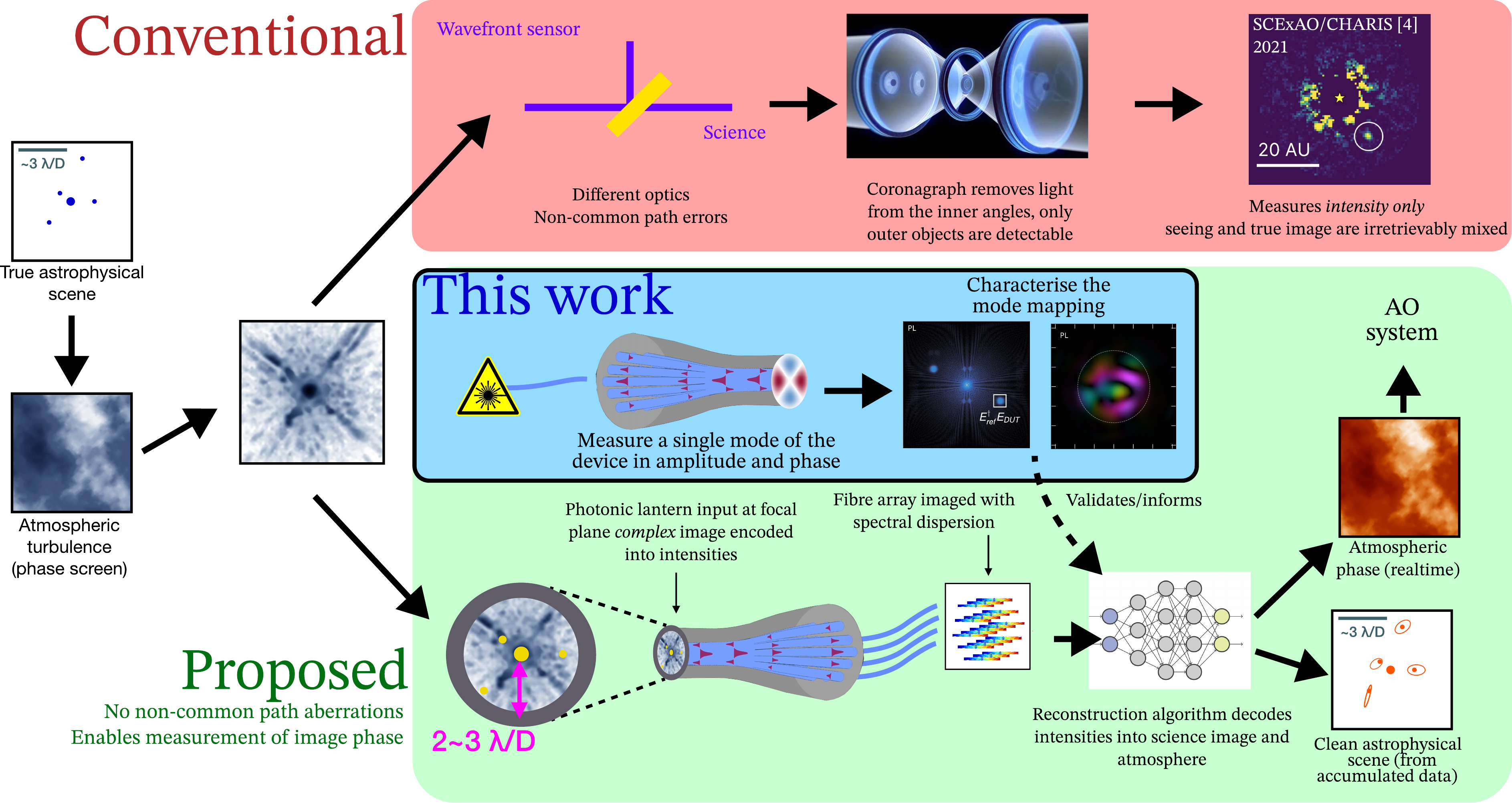}
    \caption{\textbf{Importance of empirical characterisation in an astrophotonics application.} In astronomical imaging, the true scene is distorted by atmospheric turbulence. Conventional direct imaging approaches involve separate wavefront sensing/corrections to the science signal and removing starlight using a coronagraph, which also removes information about the region closest to the star (result image from \cite{chilcote_scexaocharis_2021}). The proposed approach instead uses a photonic lantern to encode the complex electric field into intensities. A reconstruction algorithm, validated and informed by a separate characterisation process, is then able to infer images from the innermost angles of the system as well as provide real-time feedback to the adaptive optics (AO) system to correct for the atmosphere. Coronagraph image credit: \href{https://science.nasa.gov/mission/roman-space-telescope/coronagraph/}{NASA}.}
    \label{fig:intro_overview}
\end{figure}

\autoref{fig:intro_overview} visually depicts one of many possible avenues for astrophotnics in a direct imaging context. Typically, chronographic systems measure intensity only images of a scene where the starlight is blocked. As the wavefront sensor is split into a different set of optics, non-common path errors are present, which cause speckles in the final image. This forms an ambiguity when looking at intensity only imagery -- a persistent, bright pixel may  be a true companion or the result of imperfect correction. This work forms part of an alternate architecture, proposing the use of a photonic lantern to encode the complex electric field at the image plane into the intensities at single-mode outputs. A reconstruction algorithm then jointly estimates the astrophysical scene and the state of the wavefront. This estimator can exploit both the amplitude, phase and mutual coherence at the image plane, yielding richer information than intensities only. For such an architecture to function and make convincing detections, knowledge of the mapping between the inputs and outputs of the photonic lantern is critical and will be used to validate/inform the reconstruction method. 

Previous work has only just begun to measure this mapping, typically using intensity only measurements at a few wavelengths. There has recently been a push to characterise these devices in electric field~\cite{xin_laboratory_2024,Taras2026OEIllum}. In this work we analyse the mode conversion properties of photonic lanterns as measured by digital off-axis holography. Specifically, we explore the orthogonality of the principal modes (the modes corresponding to the excitation of one of the single mode ports) from a previously published transfer matrix. Furthermore, we explore (in simulation) a method of measuring the relative phases between modes of different ports.

\section{EMPIRICAL VERIFICATION OF MODE ORTHOGONALITY}

\begin{figure}[H]
    \centering
    \includegraphics[width=0.8\linewidth]{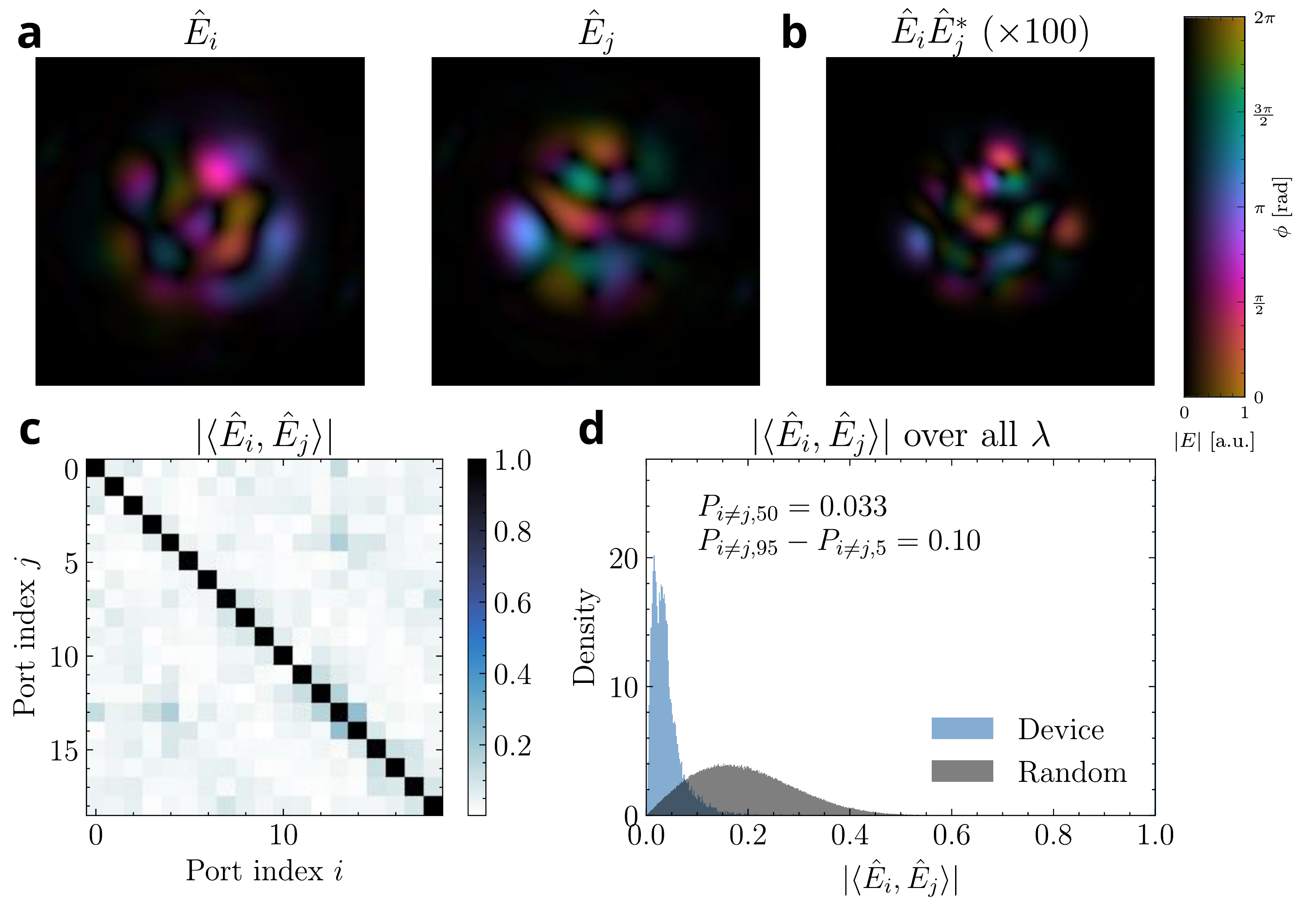}
    \caption{\textbf{Orthogonality of photonic lantern principal modes}. \textbf{a} Normalised electric fields for two different cores at the same wavelength and polarisation. \textbf{b} The complex product produces a result that has a mix of values and phases. The inner product is the complex sum of this image. \textbf{c} The inner product for all combinations of ports measures the orthogonality of the modes. \textbf{d} The  distribution from repeating this over all wavelengths measures the orthogonality of the photonic lantern principal modes. The distribution is summarised with a median $P_{50}$ and 5-95 percentile range $P_{95}-P_5$ of overlap integrals that are taken over all combinations of different ports ($i\neq j$). For comparison, a distribution from random normalised vectors (black) is also shown. }
    \label{fig:field_overlaps}
\end{figure}

One important property of mode sorting devices is the orthogonality of the supported modes. In applications where the modal mapping is inverted to infer the electric field at the multimode end, mode sorters with orthogonal modes will provide better conditioned solvers. We further examine the data product produced in previous work~\cite{Taras2026OEIllum} to verify the orthogonality of principal modes in the photonic lantern. The results are shown in \autoref{fig:field_overlaps}. For each combination of normalised electric fields from two ports, the overlap integral is computed, which corresponds to the cosine of the angle between the two complex vectors that represent each field. The majority of cases approach zero, which indicates orthogonality. \autoref{fig:field_overlaps}\textbf{c} illustrates that the less orthogonal principal modes tend to be those towards the outside of the \gls{MCF} (higher port index) and tend to be from neighbouring ports. Statistics are also taken over all wavelengths. To demonstrate that this is not an effect of the dimension of the space (noting that random vectors in large directions are nearly orthogonal), we compute the distribution for randomly draw normalised vectors in 23 dimensions (equal to the number of modes that are supported at the multimode end), which has a much larger value. An adiabatic photonic lantern must have orthogonal principal modes, and hence the result is consistent with a near-adiabatic device, corroborating the high throughput measurements made after manufacture. 
A relatively simple summary of a device's orthogonality is proposed: the median and 5-95 percentile range of the overlap integral over all different port combinations and all wavelengths. For this device, these values are 3.3\% and 10\% respectively. 

\section{TOWARDS RELATIVE PRINCIPAL MODE PHASE MEASUREMENTS}\label{sec:sim_rel_phase}

The previous measurements from the digital off-axis holographic testbed are limited in that the phase between different principal modes cannot be constrained. This is for two reasons. First, the phase is not constant due to the drifts from testbed instabilities, and measurements at different times confound the effect of these drifts with genuine differences between ports. Secondly, to excite different ports (with neighbours spaced at $60\,\mu$m in the \gls{MCF}) with no more than $\pi/10$ radians error requires tolerances of no more than $\sim0.1^\circ$ in relative tip/tilt between the two \fib{} alignment stages. Whilst it is possible to actuate the stages this finely, it is unclear how to measure this angle with sufficient accuracy.  

This limitation has no consequences when the device is used with intensity measurements of the single-mode end (such as in wavefront sensing), but matters when recombining different ports~\cite{kim_coherent_2024} or using the device for beam shaping. In this section, we use simulation to explore a remedy to this limitation by collecting data when multiple ports are excited simultaneously. The simulation used in this section does not use the dataset above; instead a random, asymmetric, unitary transfer matrix is assumed. We expect the conclusions to hold in either case. 

\begin{figure}[h]
    \centering
    \includegraphics[width=0.9\linewidth]{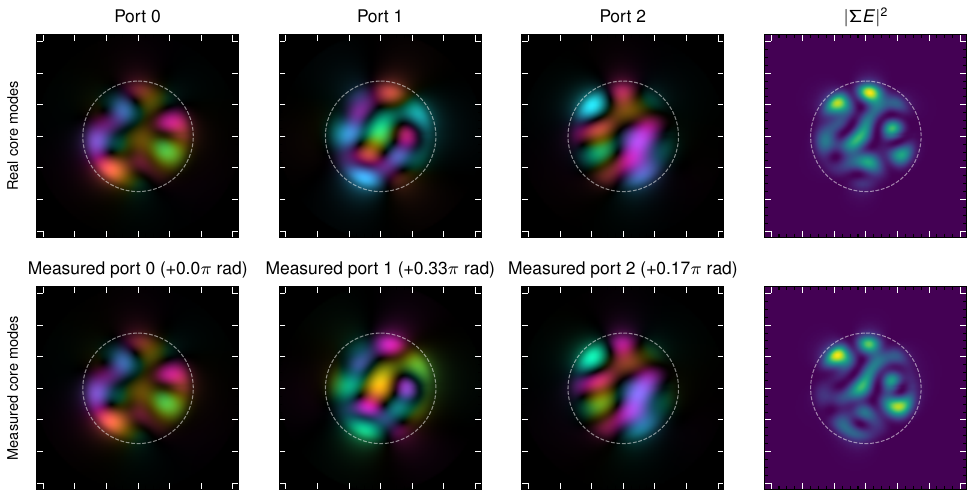}
    \caption[Simulated superpositions of port outputs reveal relative phase dependent behaviour]{\textbf{Simulated superpositions of port outputs reveal relative phase dependent behaviour.} Simulating three ports of a 19 port photonic lantern and applying a global phase offset to all ports relative to port 0 (bottom row) from the true, relative port mapping (top row). The intensity (right column) of the superposition of the three fields is different in each case, providing a means to infer relative phase at a sub-radian level.}
    \label{fig:3port_extra_measurement}
\end{figure}

The key insight leveraged is that it is possible to inject multiple ports at once in a phase deterministic way -- by pulling the \fib{} injection stage back and turning up the laser power, a superposition of the known principal modes for multiple ports is excited. As the number of pixels is larger than the number of unknown, spatially constant phases between ports, there could be enough information to disentangle the phases.

To demonstrate this in a minimal case, we start with an example of 3 ports, where we will recover the relative phases (i.e. 2 phases total). \autoref{fig:3port_extra_measurement} shows the results of a simulation of 3 ports of a photonic lantern, with real modes and relative phases (top row) compared with the same modes but with a phase offset relative to port 0 (bottom row). The intensity of the sum of the fields $|\sum E|^2$ is visually different and hence a single measurement of the mode intensity when illuminating 3 ports provides information on the relative phase between ports. 

\begin{figure}[h]
    \centering
    \includegraphics[width=0.5\linewidth]{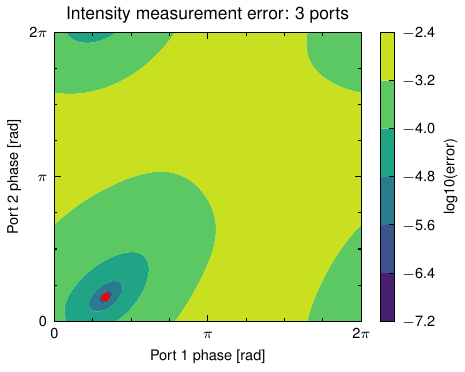}
    \caption[Contour plot of loss landscape for relative phase solving]{\textbf{Contour plot of loss landscape for relative phase solving.} The sum of squared errors {$(I_{\text{meas}} - |\sum_i E_i \exp(i\phi_i)|^2)^2$} for different port phases $\phi_1$ and $\phi_2$. The loss landscape is smooth and shows a unique minimum (red dot) when taking into account phase wrapping (and solving on a torus). }
    \label{fig:3port_phase_guess_error}
\end{figure}

The next questions are: given this additional intensity measurement, is there enough information to constrain the relative phases? If so, is such a solution unique? \autoref{fig:3port_phase_guess_error} illustrates the loss landscape to answer these questions. Each measured electric field from the port $E_i$ is multiplied by a global phase $\phi_i$, and the resulting photometric error between the measured intensity and the predicted intensity is shown. Even for complicated mode fields, the loss space is smooth and has a unique minimum. There is also a non-zero covariance between solutions, with the contour profile not aligned with either axis.

\begin{figure}[b]
    \centering
    \includegraphics[width=0.6\linewidth]{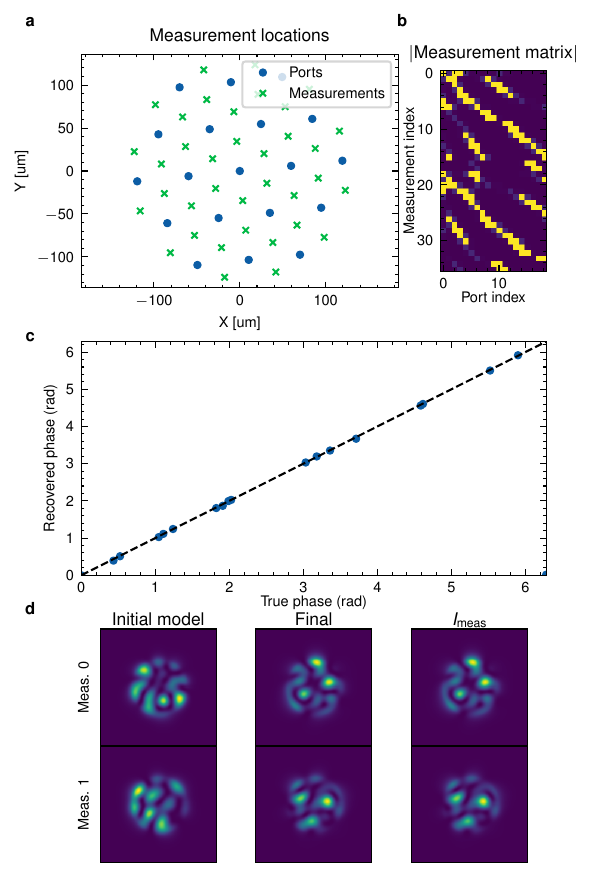}
    \caption[Simulated recovery of relative phase between principal modes]{\textbf{Simulated recovery of relative phase between principal modes.} \textbf{a} Capturing additional measurements for a 19 port lantern involves sampling all possible combinations of 2 or 3 ports. \textbf{b} Absolute value of injected electric field into each port for each measurement. The injection is such that superpositions of different ports are excited at each measurement, with a known (zero) differential phase. The model of a Gaussian beam has small leakage in other ports. \textbf{c} The recovered phase vs the true phase, showing there is enough information to recover the relative phase between all ports. \textbf{d} Two measurement examples (rows), showing the initial model (left), final model (centre) and the measured intensity (right), showing good agreement as well as the diversity of combined mode fields possible which enables precise phase measurements.  }
    \label{fig:all_phase_recovery}
\end{figure}

Finally, the simulation is generalised to solve for all relative phases, with many measurements. For the hexagonal grid, the optimal injection points are equidistant to the ports. In this part of the simulation, we include the Gaussian beam profile when illuminating ports rather than the idealised, pure 3 port problem shown above. We do, however, assume that the phase is flat along the wavefront (which isn't strictly true for ports outside the nearest three, however the relation is deterministic and hence wouldn't affect the conclusions). Future work could determine how well this applies to real data.

\autoref{fig:all_phase_recovery} reports the success of the recovery. Additional simulated measurements are taken from the locations shown by taking the superposition of the true modes formed by each port. The measurement matrix is known to the solver, and can be used to infer the relative phases. This optimisation problem is solved using the \gls{BFGS} algorithm (implemented in \texttt{scipy}~\cite{virtanen2020scipy}) and initialise from all phases as zero. The recovered phase is consistent with the true phase despite the distant initialisation in some places, indicating that the insights from \autoref{fig:3port_phase_guess_error} generalise to higher dimensions. The predictions of the model fit the additional measurements well. 

The results presented in this section show that the problem of relative phases between principal modes is solvable with a testbed similar to ours, assuming practical matters like temporal stability and tip/tilt tolerances can be met. While the principle modes do contain enough information to constrain this information, additional work is needed to verify the precision that this can be achieved in the presence of noise and possible cladding modes.

\section{CONCLUSION}
To conclude, we have shown that the principal modes of a photonic lantern (those formed by exciting only one single mode port) are very close to orthogonal. We quantify the degree of orthogonality with summary statistics of overlap integrals between modes, and hypothesise that this provides a good proxy for the quality of a manufactured device for inference problems such as wavefront sensing. We have also described and simulated a process enables future characterisation datasets to recover the phase between principal modes with additional measurements. 
Future work involves applying this to multiple photonic lanterns to asses repeatability of manufacturing methods.

\acknowledgments 

We would like to acknowledge the helpful discussions and insight from all our collaborators on the photonic lantern sensing effort, including Michael Fitzgerald, Olivier Guyon, Nemanja Jovanovic, Yoo Jung Kim, Manon Lallement, Jonathan Lin, Julien Lozi, Sebastien Vievard, Yinzi Xin, and others from California Institute of Technology, Univ. of California Los Angeles, Univ. of California Irvine, and the SCExAO team at the Subaru Telescope, NAOJ. 

B. Norris is the recipient of an Australian Research Council Discovery Early Career Award (Grant No. DE210100953) funded by the Australian Government. S. Leon-Saval and C. Betters acknowledge support by the Air Force Office of Scientific Research under award number FA2386-23-1-4108.

Microsoft Copilot was used in developing the software for this work, and Google Gemini was used in editing the final text after writing. We acknowledge support from Astralis--Australia’s optical astronomy instrumentation consortium--through the Australian Government's National Collaborative Research Infrastructure Strategy (NCRIS) program. 

We would like to acknowledge the Gadigal People of the Eora nation, the traditional owners of the land on which most of this work was completed. 

These proceedings contain content from a masters thesis\cite{taras2025illuminating} that has otherwise not been published elsewhere.

\bibliography{report} 
\bibliographystyle{spiebib} 

\end{document}